\documentclass[aps,twocolumn,groupedaddress,showpacs]{revtex4}
\usepackage{graphicx}
\begin{document}
\title{Dynamical properties of $S=1$ bond-alternating Heisenberg chains in transverse magnetic fields}
\author{Takahumi Suzuki and Sei-ichiro Suga}
\affiliation{Department of Applied Physics, Osaka University, Suita, Osaka 565-0871, Japan}
\date{\today}
\begin{abstract}
We calculate dynamical structure factors of the $S=1$ bond-alternating Heisenberg chain with a single-ion anisotropy in transverse magnetic fields, using a continued fraction method based on the Lanczos algorithm. In the Haldane-gap phase and the dimer phase, dynamical structure factors show characteristic field dependence. Possible interpretations are discussed in viewpoint of the field dependence of the excitation continuum. 
The numerical results are in qualitative agreement with recent results for inelastic neutron-scattering experiments on the $S=1$ bond-alternating Heisenberg-chain compound $\rm{Ni(C_{9}D_{24}N_{4})(NO_{2})ClO_{4}}$ and the $S=1$ Haldane-gap compound $\rm{Ni(C_{5}D_{14}N_{2})_{2}N_{3}(PF_{6})}$ in transverse magnetic fields. 
\end{abstract}
\maketitle
\section{INTRODUCTION}
$S=1$ bond-alternating Heisenberg chains have attracted a great amount of attention both theoretically and experimentally. The system shows a quantum phase transition between the Haldane-gap phase and the dimer phase depending on the bond-alternating ratio \cite{ah}. 
Thermodynamic properties at the gapless point \cite{Hagi3,kohno} and in the dimer phase \cite{nakano,narumi1,tate,narumi2} were studied experimentally. The results were compared with theoretical results and good agreement was obtained \cite{Hagi3,kohno,nakano,narumi1,narumi2}. 
The elementary excitation was investigated by quantum Monte Carlo \cite{yamamoto} and exact diagonalization \cite{totsuka} methods. In the Haldane-gap phase close to the uniform chain, the low-lying excitations are expected to be scattering states of the domain walls in the hidden antiferromagnetic ordering \cite{yamamoto}. In the vicinity of the isolated dimer system, the low-lying excitations are well described by the $S=1$ magnon \cite{totsuka}. 
Dynamical structure factors in the Haldane-gap and dimer phases were calculated by the exact diagonalization method \cite{suzuki}. In both phases, the distributions of the scattering intensity of the one-magnon mode and the excitation continuum were investigated.

It is known that a feature of the Haldane-gap phase appears in transverse magnetic fields. In fact, the energy spectrum in transverse fields was measured for the anisotropic Haldane-gap compound $\rm{Ni(C_{2}D_{8}N_{8})_{2}(NO_{2})ClO_{4}}$ (abbreviated to NENP) by inelastic neutron-scattering experiments \cite{Regnault} and the satisfying agreement between the theoretical \cite{Tasaki} and experimental results was achieved. 
Many $S=1$ bond-alternating Heisenberg-chains compounds possess noticeable single-ion anisotropy. Characteristic spin dynamics in the Haldane-gap and dimer phases may thus emerge in transverse magnetic fields. However, little knowledge had been obtained about dynamical properties of $S=1$ bond-alternating Heisenberg chains in transverse magnetic fields.

 Quite recently, inelastic neutron-scattering experiments on the dimer-phase compound $\rm{Ni(C_{9}D_{24}N_{4})(NO_{2})ClO_{4}}$ (abbreviated to NTENP) in transverse magnetic fields ($H$) revealed fascinating aspects of dynamical properties \cite{Hag1}.  
In $H=0$, the triplet magnon excitation is lifted into the lower $S^{z}=\pm1$ branch and the higher $S^{z}=0$ branch owing to a single-ion anisotropy. 
The observed scattering intensity of the higher $S^{z}=0$ branch was anomalously weak at $q=\pi$ as compared with that in the anisotropic Haldane-gap compound $\rm{Ni(C_{5}D_{14}N_{2})_{2}N_{3}(PF_{6})}$ (abbreviated to NDMAP), where the wave number is represented in the extended zone scheme. 
In $H \neq 0$, the degenerate lower branch is lifted. As $H$ increases, the excitation energy of the highest (lowest) branch increased (decreased), while that of the middle branch was almost independent of $H$. 
In NTENP, the intensity of the highest branch decreased rapidly with increasing $H$ and disappeared at $H \sim 0.35H_{\rm c}$, whereas the lower two branches survived up to $H=H_{\rm c}$. Note that $H_{\rm c}$ is the critical field where the excitation gap closes. 
In $H>H_{\rm c}$, there appeared only one gapped branch at $q=\pi$ which continued from the middle branch in $H<H_{\rm c}$. 
These findings are in contrast with those observed in NENP \cite{Regnault} and NDMAP \cite{Zhe1} under transverse magnetic fields: In $H \leq H_{\rm c}$ three branches showed conspicuous intensity up to $H=H_{\rm c}$ and in $H>H_{\rm c}$ three gapped branches were observed. 
It is desirable to clarify the origin of the different spin dynamics between these systems.

In this paper, we calculate the dynamical structure factor (DSF) of the $S=1$ bond-alternating Heisenberg chain with a single-ion anisotropy in transverse magnetic fields, using a continued-fraction method based on the Lanczos algorithm \cite{GB}. By shifting the bond-alternating ratio systematically, dynamical properties in the Haldane-gap phase and the dimer phase are investigated. 
In Sec. II, we briefly summarize the method for the numerical calculation. In Sec. III, we show the results for the DSF, turning our attention to the behavior at $q=\pi$. 
In Sec. IV, the characteristic field dependence in the Haldane-gap phase and dimer phase is discussed in viewpoint of the field dependence of the excitation continuum. The origin of different spin dynamics observed in NTENP and NDMAP is also discussed. Sec. V is devoted to the summary.

\section{Model and Method} 
Let us consider the $S=1$ bond-alternating Heisenberg chain with a single-ion anisotropy in transverse magnetic fields described by the following Hamiltonian,
\begin{eqnarray}
\mathcal{H} &=& J\sum_{i} \Big(\mathbf{S}_{2i-1} \cdot \mathbf{S}_{2i} + 
            \alpha \mathbf{S}_{2i} \cdot \mathbf{S}_{2i+1}\Big) \nonumber\\
  &+& D \sum_{i} (S^{z}_{i})^{2} - g\mu_{B}H \sum_{i} S_{i}^{x},  
\label{Ham1}
\end{eqnarray}
where $J>0$ and $\alpha$ is a bond-alternating ratio. 
The periodic boundary condition is applied. 
The DFS can be expressed as \cite{GB}
\begin{eqnarray}
S^{\mu\nu}(q,\omega)
&=& -\frac{1}{\pi} {\rm Im} 
\langle\Psi_{g}|S^{\mu}_{q}\frac{1}{z-\mathcal{H}}S^{\nu}_{q}|\Psi_{g}\rangle  
\nonumber \\
&=& S^{\mu\nu}(q) C^{\mu\nu}(q,\omega) \hspace{5mm} (\mu=x,y,z) ,
\label{Ham2}
\end{eqnarray}
where $| \Psi_{g} \rangle$ is the eigenstate of $\mathcal{H}$ with the lowest eigenvalue $E_{g}$, $S^{\mu}_q=(1/\sqrt{N})\sum_{j}e^{iqj}S^{\mu}_{j}$ with $N$ being the total number of spins, and $z=\hbar\omega+i\eta+E_{g}$. The lattice constant between neighboring two spins is set to unity. Therefore, $q=0.5\pi$ corresponds to the boundary of the Brillouin zone in case of $\alpha\ne1$.  
We set $g\mu_{B}=1$ and $\hbar=1$. The energy is measured in units of $J$.

In the expression (\ref{Ham2}), $S^{\mu\nu}(q)$ is the static structure factor and $C^{\mu\nu}(q,\omega)$ is represented in the form of the continued fraction, which can be calculated numerically by Lanczos algorithm. The total contribution of $C^{\mu\nu}(q,\omega)$ for a fixed $q$ is normalized to unity, because the sum rule, $S^{\mu\nu}(q)=\int^{\infty}_{0} {\rm d}\omega S^{\mu\nu}(q,\omega)$, has to be satisfied. Instead of taking $\eta\to+0$, we set $\eta = 1.0 \times 10^{-2}$. In finite-size systems, therefore, $C^{\mu\nu}(q,\omega)$ consists of a finite number of Lorentzians. The integrated value of each Lorentzian with respect to $\omega$ in given $q$ is called as the residue \cite{Tak1}. 
To discuss whether an excited state forms the isolated mode or the excitation continuum in the thermodynamic limit, the criterion proposed by Takahashi \cite{Tak1} may be effective. 
According to this criterion, the residue of an excited state in the continuum tends to decrease with increasing the system size $N$, while the residue for an isolated mode hardly depends on $N$. This method was successfully applied \cite{Tak1} to explain spin dynamics of NENP observed by inelastic neutron-scattering measurements \cite{Ma}.

Adopting the parameter sets $(\alpha,D)=(1.0,0.25), (0.8,0.2), (0.45,0.25)$, and $(0.25, 0.08)$, we calculate the DFS up to $N=20$ spin systems in $H=0$ and $N=16$ spin systems in $H \ne 0$. The parameter sets $(\alpha,D)=(1.0,0.25), (0.45,0.25)$, and $(0.25, 0.08)$ describe NDMAP \cite{Zhe2}, NTENP \cite{narumi1}, and the dimer-phase compound NMAOP \cite{narumi3}, respectively.
According to the phase diagram \cite{tone}, the parameter set $(\alpha,D)=(0.8,0.2)$ belongs to the Haldane-gap phase and is located nearer the gapless line than NDMAP \cite{com}. In the dimer phase, NTENP is situated nearer the gapless line than NMAOP. 
In the next section, we show the results for $S^{zz}(q,\omega)$ and $S^{\perp}(q,\omega) \equiv S^{xx}(q,\omega)+S^{yy}(q,\omega)$.

\section{Numerical Results} 
\subsection{$H=0$} 
\begin{figure}[thb]
\begin{center}
\includegraphics[trim=0mm 0mm 0mm 0mm,clip,scale=0.12]{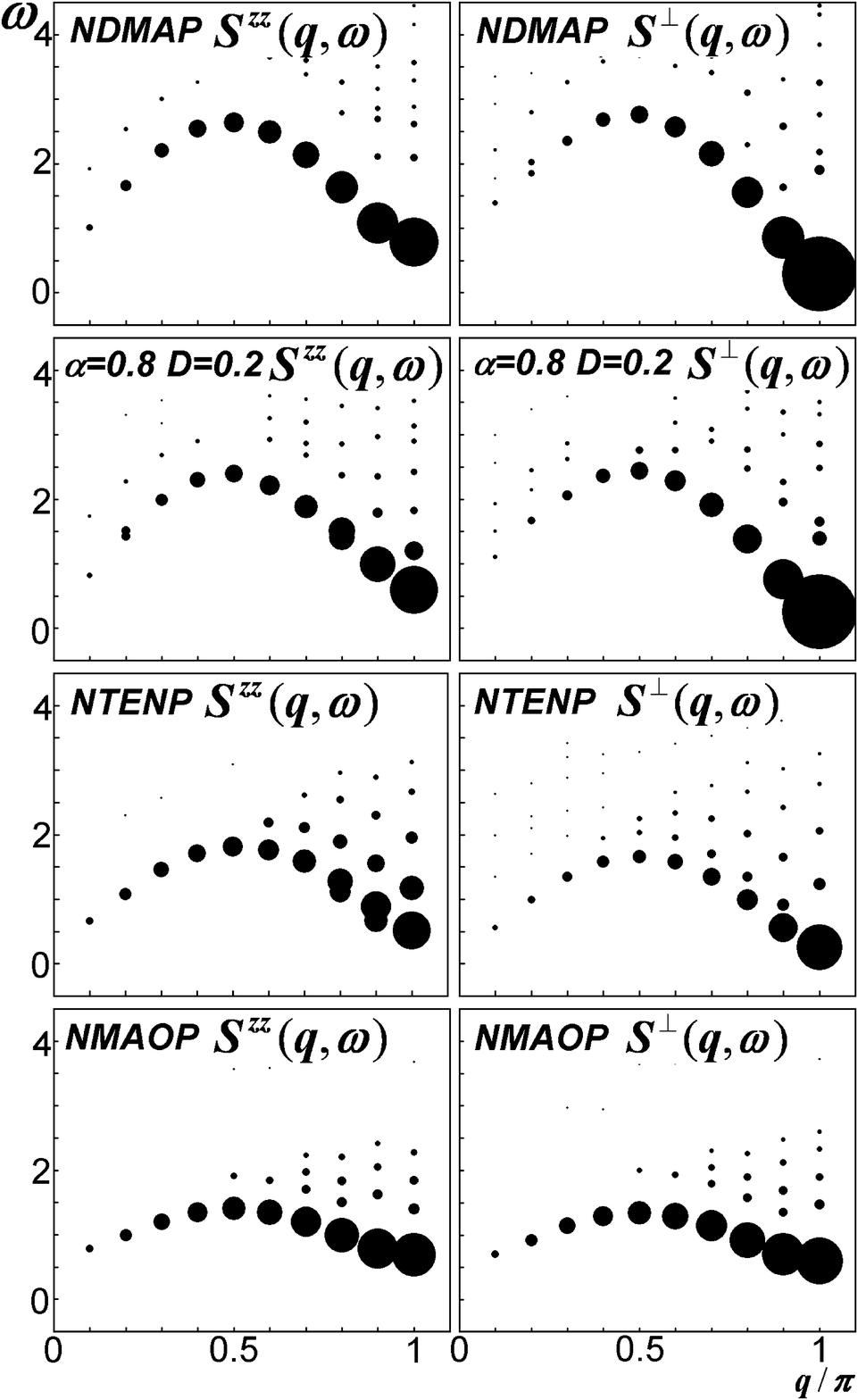}
\end{center}
\caption{$S^{zz}(q,\omega)$ and $S^{\perp}(q,\omega)$ at $H=0$ for $N=20$. The area of the circle is proportional to the scattering intensity. 
}
\label{fig1}
\end{figure}

The DSF's in $H=0$ are shown in Fig. \ref{fig1}. Except for NDMAP, the results are shown in the extended zone scheme. The area of the circle is proportional to the scattering intensity. We turn our attention to the behavior at $q=\pi$. 
In $(\alpha,D)=(0.8,0.2)$ and NTENP, the next-lowest-excited state in $S^{zz}(\pi,\omega)$ with relatively large intensity is located close to the lowest excited state, while in $S^{\perp}(\pi,\omega)$ the next-lowest-excited state lies well above the lowest excited state. 
In NDMAP and NMAOP, there appear appreciable gaps between them in both $S^{zz}(\pi,\omega)$ and $S^{\perp}(\pi,\omega)$. 
\begin{figure}[thb]
\begin{center}
\includegraphics[trim=0mm 0mm 0mm 0mm,clip,scale=0.12]{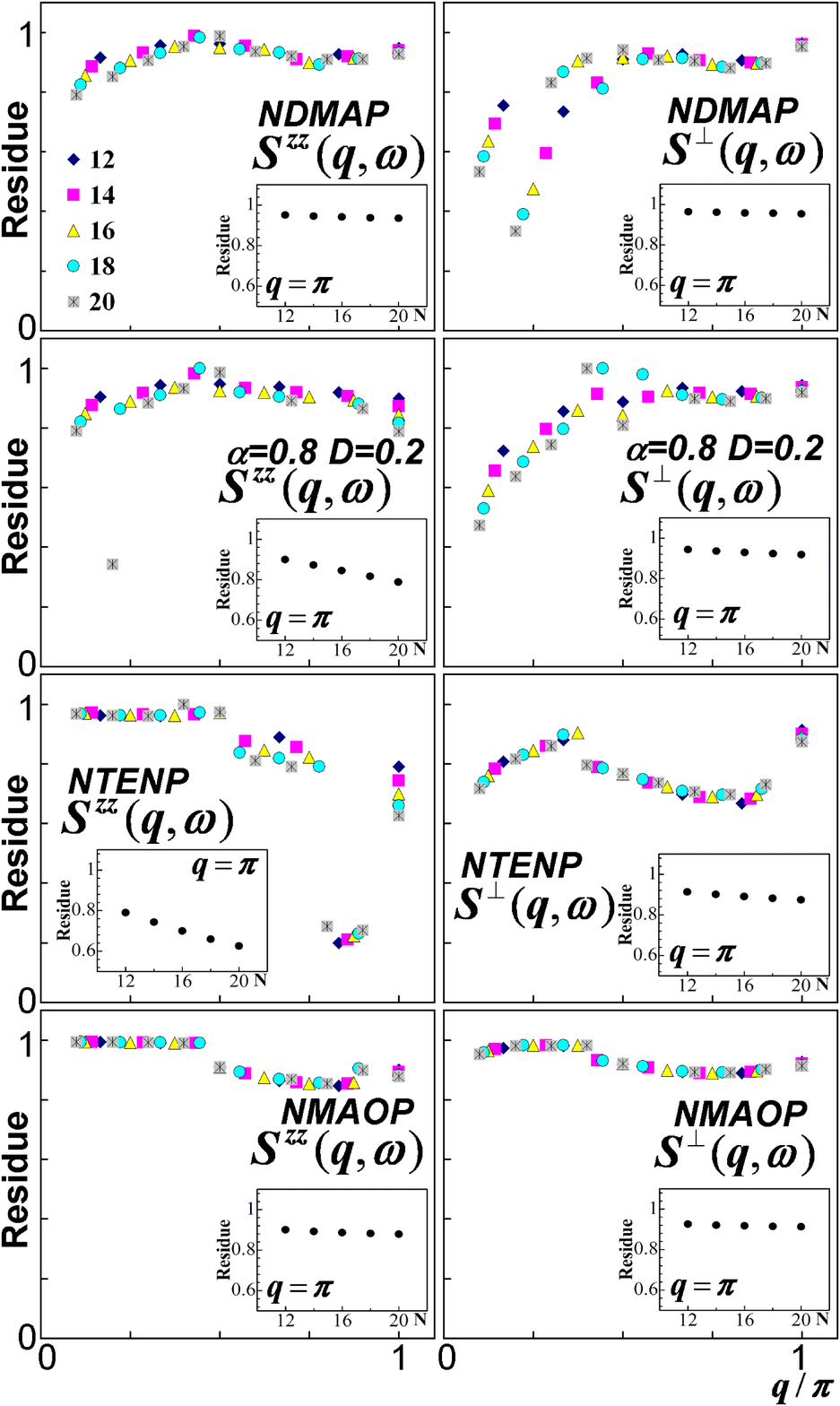}
\end{center}
\caption{The size dependence of the residue in $H=0$. 
Insets: The $N$ dependence of the residue at $q=\pi$. 
}
\label{fig2}
\end{figure}

In Fig. \ref{fig2}, we show the finite-size effects of the residues of the lowest excited states in $S^{zz}(q,\omega)$ and $S^{\perp}(q,\omega)$. 
As shown in the insets, the residues of $S^{zz}(\pi,\omega)$ and $S^{\perp}(\pi,\omega)$ in NDMAP and NMAOP show little size dependence, indicating that their lowest excited states form the isolated modes. 
In $(\alpha,D)=(0.8,0.2)$ and NTENP, on the contrary, the residues in $S^{zz}(\pi,\omega)$ decrease with increasing $N$, while the residues in $S^{\perp}(\pi,\omega)$ scarcely depend on $N$. In spite of such appreciable size dependence, we have to be careful to decide the characteristic of the lowest excited state of $S^{zz}(\pi,\omega)$ in $(\alpha,D)=(0.8,0.2)$ and NTENP. 
Since their next-lowest-excited states are located close to the lowest excited states as shown in Fig. \ref{fig1}, the residues of the lowest excited states may be suffering from the higher energy states.

\begin{figure}[thb]
\begin{center}
\includegraphics[trim=0mm 0mm 0mm 0mm,clip,scale=0.24]{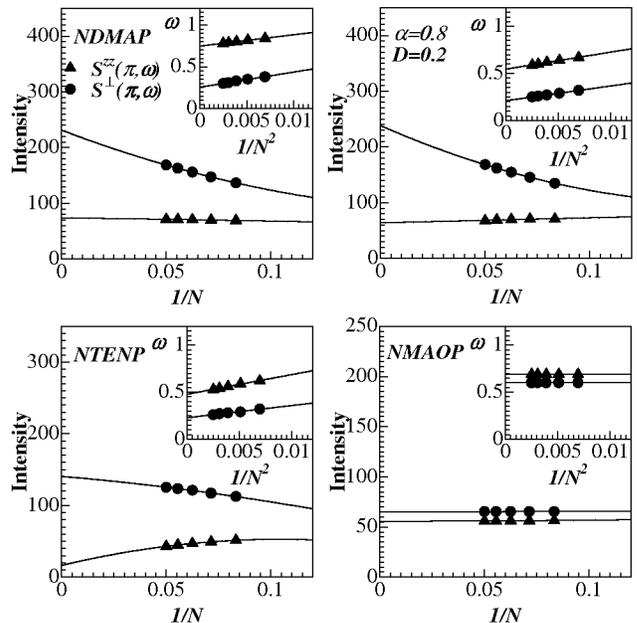}
\end{center}
\caption{The intensity and the excitation energy (insets) of the lower excited states extrapolated to $N \rightarrow \infty$ in $H=0$.}
\label{fig3}
\end{figure}
The excitation energy and the intensity of the watched lower excited states are extrapolated to $N \rightarrow \infty$. Their $N$ dependence is shown in Fig. \ref{fig3}. 
We first evaluate the ratio of the intensity of the lowest excited state in $S^{zz}(\pi,\omega)$ to that in $S^{\perp}(\pi,\omega)$ at $H=0$. The lowest excited states in $S^{zz}(\pi,\omega)$ and $S^{\perp}(\pi,\omega)$ at $H=0$ are the $S^{z}=0$ and $S^{z}=\pm1$ branches, respectively. 
The resultant ratios are $0.35$ for NDMAP, $0.29$ for $(\alpha,D)=(0.8,0.2)$, $0.11$ for NTENP, and $0.87$ for NMAOP. Our results for NDMAP and NTENP are consistent with the experimental results \cite{Hag1}.

\subsection{$H<H_{\rm c}$} 
\begin{figure}[thb]
\begin{center}
\includegraphics[trim=0mm 0mm 0mm 0mm,clip,scale=0.52]{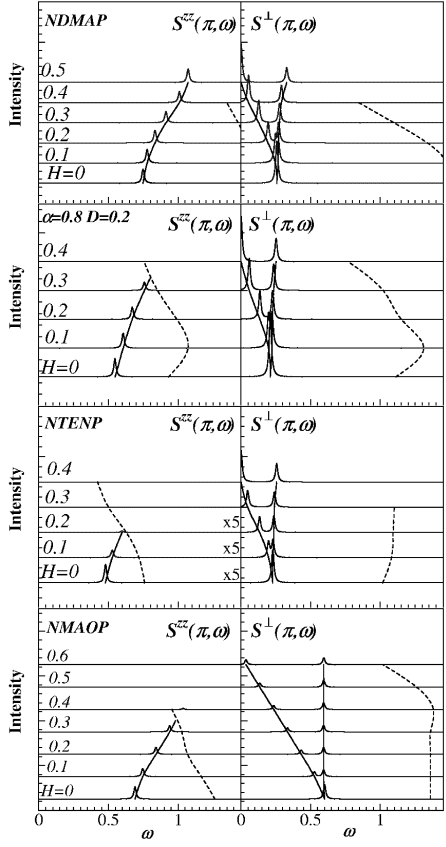}
\end{center}
\caption{The field dependence of the excitation energy and intensity for the isolated mode and the lower edge of the excitation continuum. 
In NTENP, the intensity of $S^{zz}(\pi,\omega)$ is enlarged by a factor of $5$ as compared with the intensity in the other figures. The solid and broken lines denote the excitation energies for the isolated mode and lower edge of the excitation continuum, respectively. 
The critical fields are $H_{\rm c} \sim 0.50$ for NDMAP, $H_{\rm c} \sim 0.40$ for $(\alpha,D)=(0.8,0.2)$, $H_{\rm c} \sim 0.40$ for NTENP, and $H_{\rm c} \sim 0.65$ for NMAOP. }
\label{fig4}
\end{figure}
The field dependence of the excitation energy and the intensity at $q=\pi$ is investigated in the same way. The results are summarized in Fig. \ref{fig4}. 
In magnetic fields, the lowest excited state in $S^{\perp}(\pi,\omega)$ separates into two branches. As $H$ increases, the lower branch shifts to the lower energy region and is softened at $H_{\rm c}$, which results in the quantum phase transition. The critical fields are evaluated as $H_{\rm c} \sim 0.50$ for NDMAP, $H_{\rm c} \sim 0.40$ for $(\alpha,D)=(0.8,0.2)$, $H_{\rm c} \sim 0.40$ for NTENP, and $H_{\rm c} \sim 0.65$ for NMAOP. 
Note that the gapped excitation energy is fitted well with $1/N^2$, while the softened mode is proportional to $1/N$ at $H \sim H_{\rm c}$ \cite{fss}.

We first discuss the behavior in $S^{zz}(\pi,\omega)$. As $H$ increases, the lowest excited state in $S^{zz}(\pi,\omega)$ shifts towards the higher energy region, while the next-lowest-excited state shifts to the lower energy region. 
In NDMAP, the isolated mode never meets the excitation continuum even around $H=H_{\rm c}$. Therefore, the distinct intensity of the isolated mode appears in $H \leq H_{\rm c}$. 
In $(\alpha,D)=(0.8,0.2)$, the isolated mode merges into the excitation continuum around $H=3.5(<H_{\rm c})$, where the intensity of the isolated mode becomes so weak. 
Note that there are many excited states above the lower edge of the excitation continuum. However, their intensity is so small that they are invisible in the present scale.

In NTENP and NMAOP, the intensity of the lowest excited states in $S^{zz}(\pi,\omega)$ diminishes with increasing $H$ and disappears around $H \sim 0.2$ and $\sim 0.4$, respectively, where the lowest excited states cross the excitation continua.
After the crossing, the replaced lowest excited states reach $\omega \sim 0.4$ at $H \sim H_{\rm c}$ in NTENP. 
Judging from the results for $(\alpha,D)=(0.8,0.2)$ and NTENP, at $H=0$ their lowest excited states in $S^{zz}(\pi,\omega)$ form the isolated modes.

In $S^{\perp}(\pi,\omega)$, on the other hand, the excitation continua of the four systems are located in the higher energy region up to $H \sim H_{\rm c}$. Accordingly, two isolated modes appear in $H \leq H_{\rm c}$ in $S^{\perp}(\pi,\omega)$.

\subsection{$H>H_{\rm c}$} 
\begin{figure}[thb]
\begin{center}
\includegraphics[trim=0mm 0mm 0mm 0mm,clip,scale=0.52]{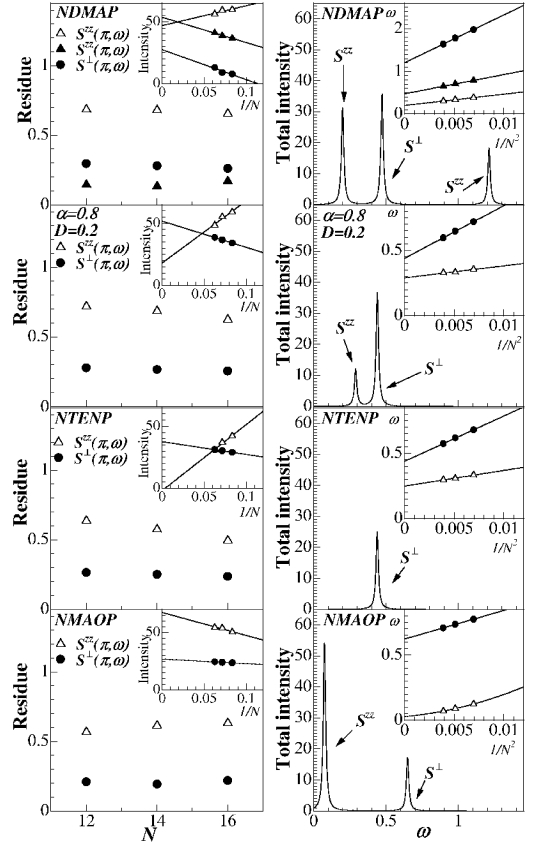}
\end{center}
\caption{The $N$ dependence of the residue at $q=\pi$ in $H>H_{\rm c}$ (left figures). $H=0.6$ for NDMAP, $H=0.5$ for $(\alpha,D)=(0.8,0.2)$, $H=0.6$ for NTENP, and $H=0.7$ for NMAOP. The extrapolated intensity and the excitation energy in $S^{zz}(\pi,\omega)$ and $S^{\perp}(\pi,\omega)$ are presented in the same figure (right figures). 
Insets: Extrapolation of the intensity (left figures) and the excitation energy (right figures) to $N \rightarrow \infty$.}
\label{fig5}
\end{figure}
We next investigate dynamical properties in $H>H_{\rm c}$ using the same method. In $H>H_{\rm c}$, the lowest-lying excitation becomes gapless, yielding the peak with the largest intensity in $S^{\perp}(\pi,0)$ irrespective of $H$. We disregard the behavior of this peak. 
In Fig. \ref{fig5}, we show the results for NDMAP at $H=0.6$, $(\alpha,D)=(0.8,0.2)$ at $H=0.5$, NTENP at $H=0.6$, and NMAOP at $H=0.7$. 
The residues of the peaks are shown in the left figures, and the extrapolated intensity and energy are presented in the right figures. 
The extrapolation of the intensity and the excitation energy are shown in the insets.

In NDMAP, two appreciable peaks appear in $S^{zz}(\pi,\omega)$, while one peak appears in $S^{\perp}(\pi,\omega)$. Their residues are almost independent of $N$ and the extrapolated intensity takes nonzero values. Therefore, these three peaks are from the isolated modes. Note that at $H=H_{\rm c}$ the lower isolated mode in $S^{zz}(\pi,\omega)$ emerges at $\omega=0$, after the lower branch in $S^{\perp}(\pi,\omega)$ becomes softened. 
In $(\alpha,D)=(0.8,0.2)$, one appreciable peak appears in $S^{zz}(\pi,\omega)$ and $S^{\perp}(\pi,\omega)$, respectively. Their residues hardly depend on $N$ and the extrapolated intensity takes nonzero values. Thus, these two peaks are from the isolated modes. 
Since the excitation continuum at $H=0.6$ is located around $\omega =0.8$, only the lower branch of $S^{zz}(\pi,\omega)$ emerges.

In NTENP, only one noticeable peak appears in $S^{\perp}(\pi,\omega)$. Its residue scarcely depends on $N$, while the residue of the lowest excited state in $S^{zz}(\pi,\omega)$ decreases with increasing $N$. The results indicate that the lowest excited state in $S^{\perp}(\pi,\omega)$ forms the isolated mode, while that in $S^{zz}(\pi,\omega)$ forms the lower edge of the excitation continuum. 
In fact, the intensity of the latter state extrapolated to $N \rightarrow \infty$ is much smaller than the extrapolated former one as shown in the inset. 
The lower edge of the excitation continuum of $S^{zz}(\pi,\omega)$ is evaluated as $\omega \sim 0.2$ at $H=0.6$, which makes the isolated mode in $S^{zz}(\pi,\omega)$ unstable in $\omega \geq 0.2$. 
On the contrary, the excitation continuum of $S^{\perp}(\pi,\omega)$ lies in $\omega \geq 1.2$. 
Accordingly, in NTENP only one peak emerges in $S^{\perp}(\pi,\omega)$.

In NMAOP, one peak appears in $S^{zz}(\pi,\omega)$ and $S^{\perp}(\pi,\omega)$, respectively. Their residues hardly depend on $N$, indicating that they are from the isolated modes. 
In $H>H_{\rm c}$, the excitation continuum in $S^{zz}(\pi,\omega)$ shifts to the low energy region. At $H=0.7$ its lower edge is located at $\omega \sim 0.2$, which is close to the isolated mode. Therefore, as $H$ increases in $H>H_{\rm c}$, the isolated mode in $S^{zz}(\pi,\omega)$ probably disappears. 
By contrast, the excitation continuum of $S^{\perp}(\pi,\omega)$ lies in $\omega \geq 1.0$, yielding the stable isolated mode in $S^{\perp}(\pi,\omega)$ even in $H>H_{\rm c}$.

\section{Discussion} 
$S^{zz}(q,\omega)$ is active on the excitation process that conserves the $S^z$ component of the ground state, while $S^{\perp}(q,\omega)$ is active on the excitation process that changes the $S^z$ component of the ground state by $\pm 1$.  In $H \geq H_{\rm c}$, the ground state with $\langle S^x \rangle \neq 0$ and $\langle S^z \rangle = 0$ takes place instead of the singlet ground state in  $H < H_{\rm c}$. In fact, we have confirmed that $\langle S^x \rangle = 0 \sim 0.2$ and $\langle S^z \rangle = O(10^{-6})$ in $H_{\rm c} \leq H \leq 0.7$ for our four parameter sets. 
In $H \geq H_{\rm c}$, therefore, the excitations to the singlet state and to the triplet state with $S^z=0$ component make main contributions to the lower and higher isolated modes of $S^{zz}(\pi,\omega)$, respectively.

\subsection{Relation to inelastic neutron-scattering experiments} 
We now discuss the observable DSF in the reduced zone scheme. 
In $(\alpha,D)=(0.8,0.2)$, the residues of $S^{zz}(q,\omega)$ and $S^{\perp}(q,\omega)$ in $q<0.4\pi$ show a typical pattern for the excitation continuum \cite{Tak1} as shown in Fig. \ref{fig2}. The excitation energies of the isolated modes at $q=\pi$ are smaller than that around $q \sim 0$ even in transverse fields. 
When $0 \leq H \leq H_{\rm c}$, therefore, in the reduced zone scheme one isolated mode appears in $S^{zz}(0,\omega)$ and two isolated modes appear in $S^{\perp}(0,\omega)$. 
In $H>H_{\rm c}$, one isolated mode emerges in $S^{zz}(0,\omega)$ and $S^{\perp}(0,\omega)$, respectively.

In NTENP and NMAOP, the excitation energies take almost the same values at the symmetric wave numbers about $q=0.5\pi$ even in transverse fields. The same feature was already known in the dimer phase for $D=0$ and $H=0$ \cite{suzuki}. 
As shown in Fig. \ref{fig2}, the residues of $S^{zz}(q,\omega)$ and $S^{\perp}(q,\omega)$ in $q<0.4\pi$ show typical behavior for the isolated mode \cite{Tak1}. Therefore, in the reduced zone scheme the lowest excited states of $S^{zz}(0,\omega)$ and $S^{\perp}(0,\omega)$ form the isolated modes in $H=0$. 
As $H$ increases, in NTENP and NMAOP the intensity of the highest branch in $S^{zz}(0,\omega)$ decreases and disappears at $H \sim 0.2$ and $0.4$, respectively. 
In $H>H_{\rm c}$, only one isolated mode of $S^{\perp}(0,\omega)$ emerges in NTENP, while the isolated modes of respective $S^{zz}(0,\omega)$ and $S^{\perp}(0,\omega)$ emerge in NMAOP.

\begin{figure}[thb]
\begin{center}
\includegraphics[trim=0mm 0mm 0mm 0mm,clip,scale=0.18]{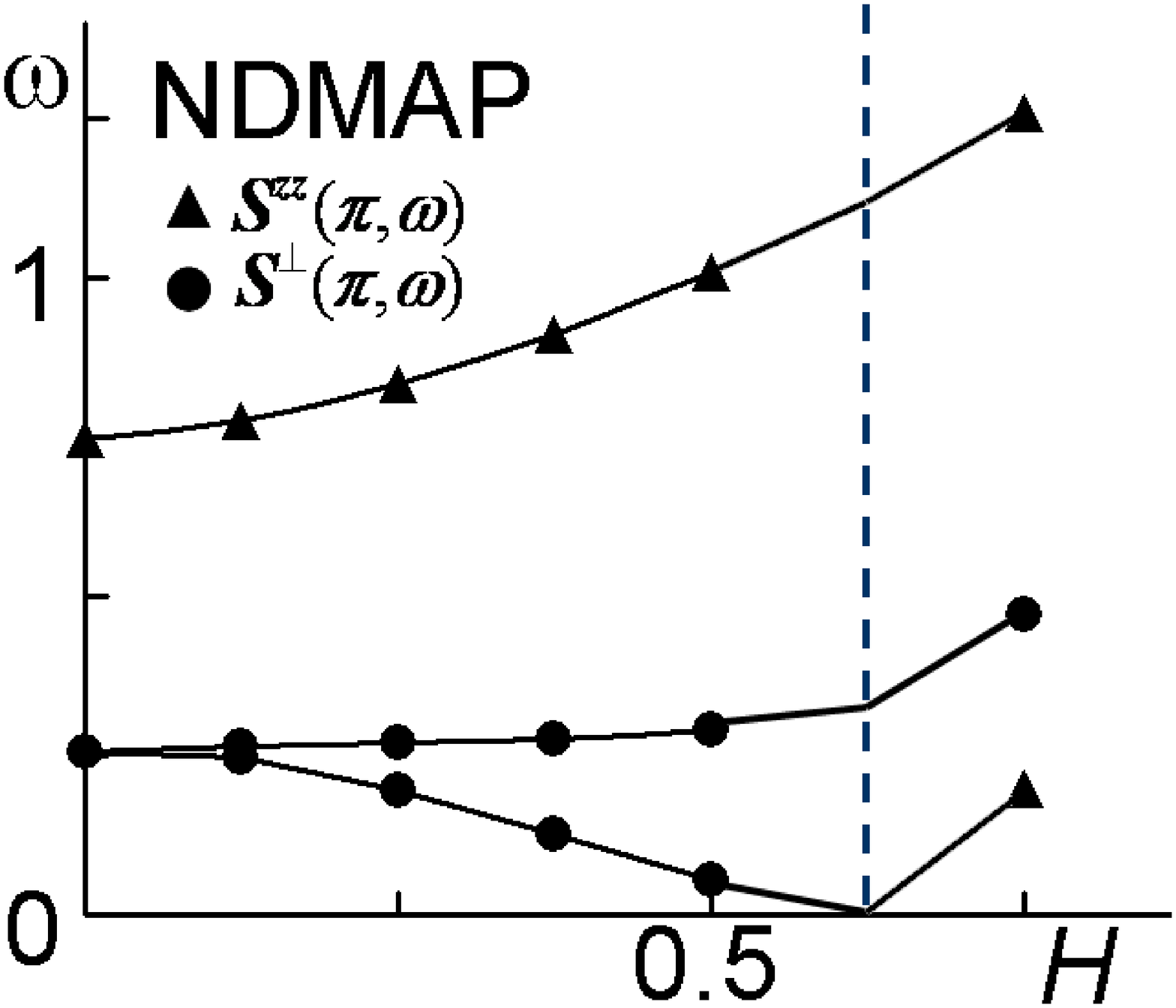}
\includegraphics[trim=0mm 0mm 0mm 0mm,clip,scale=0.18]{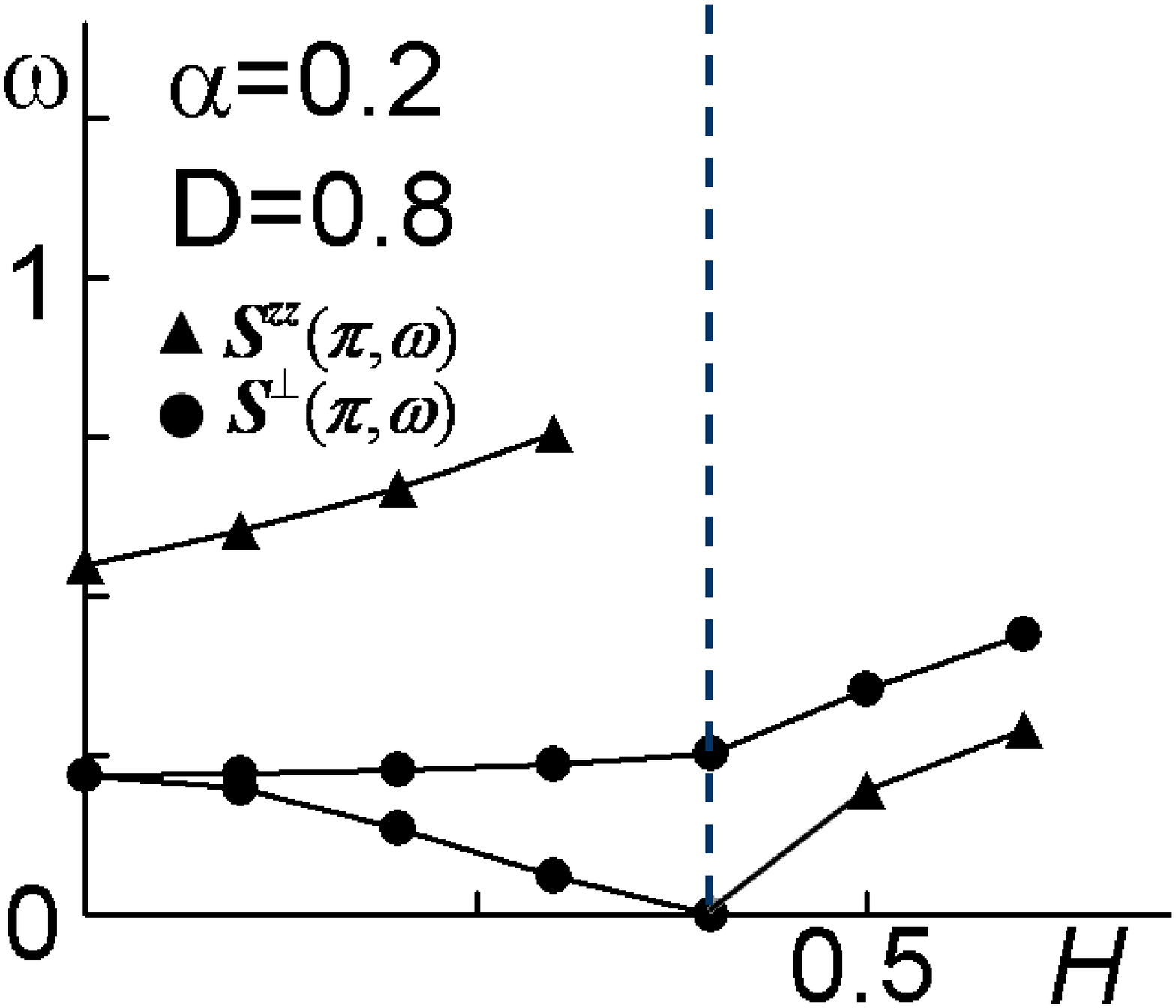}
\includegraphics[trim=0mm 0mm 0mm 0mm,clip,scale=0.18]{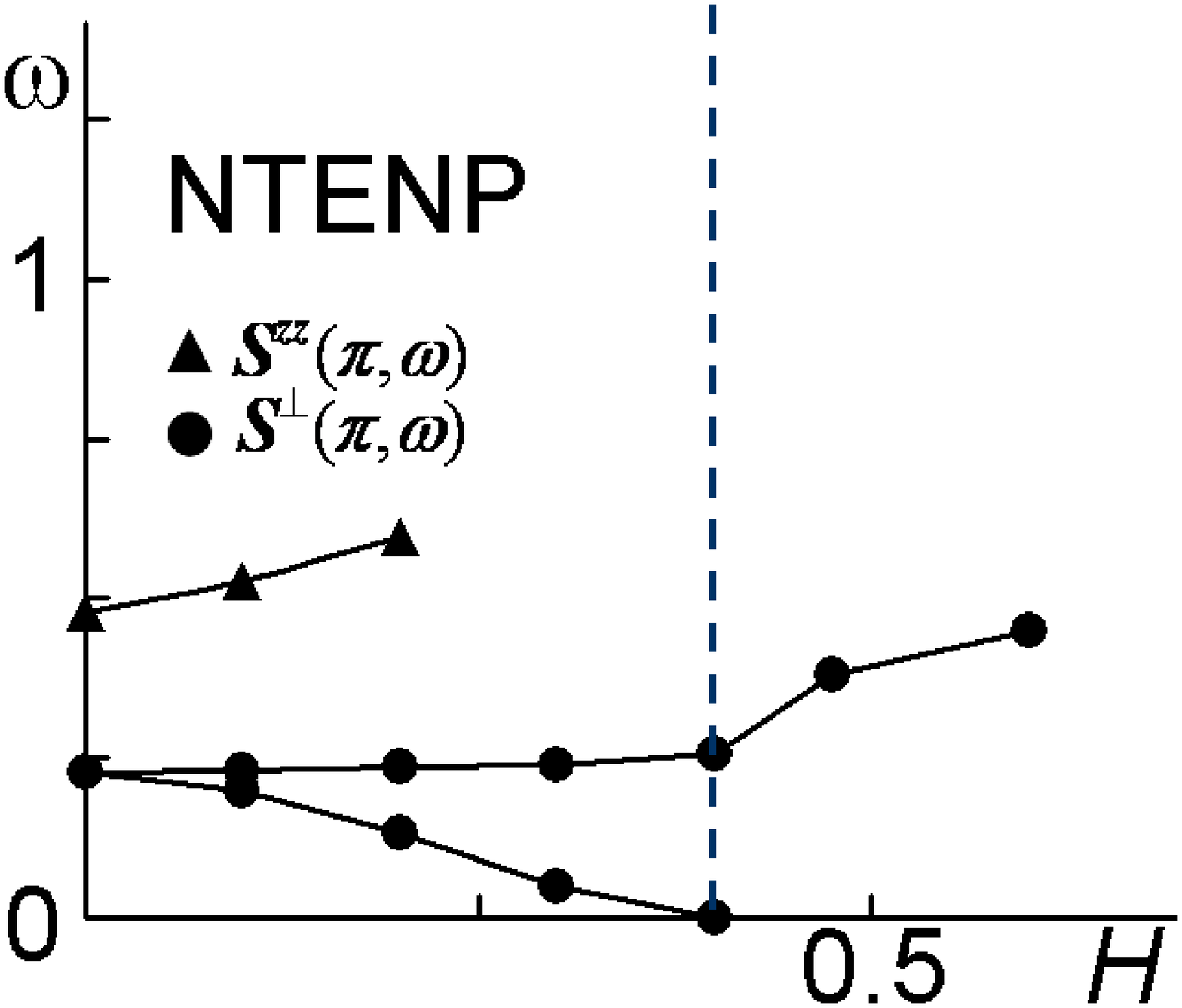}
\includegraphics[trim=0mm 0mm 0mm 0mm,clip,scale=0.18]{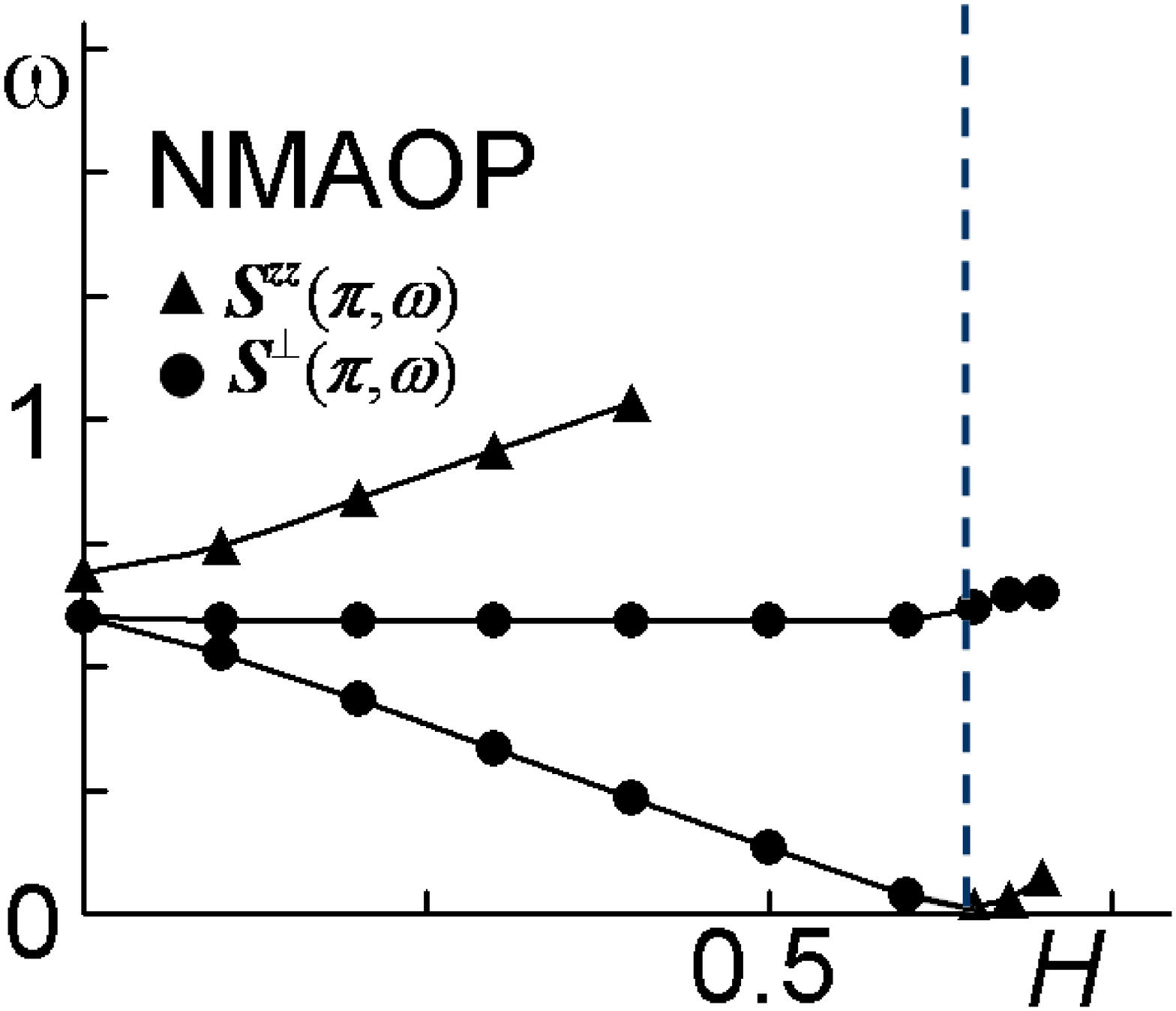}
\end{center}
\caption{The field dependence of the excitation energies at $q=\pi$. Full triangles denote $S^{zz}(\pi,\omega)$ and gray circles denote $S^{\perp}(\pi,\omega)$. On the solid and gray lines, appreciable scattering intensity appears. The broken lines for $H_{\rm c}$ are guides to eyes. }
\label{fig6}
\end{figure}
In inelastic neutron-scattering experiments on NTENP and NDMAP \cite{Hag1,Zhe1}, the DSF's were observed as a superposition of $S^{zz}(q,\omega)$ and $S^{\perp}(q,\omega)$. 
Using our numerical results, we summarize the corresponding results in Fig. \ref{fig6}. On the solid and gray lines, the isolated modes possess appreciable intensity. 
In NTENP, the highest branch disappears at $H \sim 0.5H_{\rm c}$, which is larger than the value $\sim 0.35H_{\rm c}$ estimated experimentally \cite{Hag1}. 
Above $H_{\rm c}$, only one gapped branch emerges in NTENP, whereas three gapped branches appear in NDMAP. 
The field dependence of the DSF's shown in Fig. \ref{fig6} reproduce qualitatively the experimental results for NTENP and NDMAP \cite{Hag1,Zhe1}.

In NTENP \cite{tate} and NDMAP \cite{honda}, the phase transitions to the three-dimensional ordered states were observed by specific heat measurements in transverse fields $H\geq H_{\rm c}$ at low temperatures. This phase transition is caused by the interchain interaction $J^{\prime}$. In NTENP and NDMAP, their values are estimated as $J^{\prime} \leq 3\times 10^{-3}J$ \cite{Zhe3}, and $J^{\prime} \sim 6\times 10^{-4}J$ and $J^{\prime} \leq 1\times 10^{-4}J$ depending on the direction \cite{Zhe2}, respectively. Judging from our numerical results, the interchain interactions in NTENP and NDMAP are so weak that they do not have serious effects on dynamical properties in $H>H_{\rm c}$.

\subsection{Spin dynamics in the Haldane-gap phase and the dimer phase} 
The characteristic dynamical properties are attributed to the different field dependence of the excitation continuum in $S^{zz}(\pi,\omega)$. 
As the system approaches the gapless line in both phases, the excitation continuum of $S^{zz}(\pi,\omega)$ easily shifts to the low energy region in weak fields. In the Haldane-gap phase this feature makes the highest isolated mode of $(\alpha,D)=(0.8,0.2)$ unstable and invisible in $H \geq H_{\rm c}$, whereas in NDMAP three isolated modes appear even in $H \geq H_{\rm c}$. 
When $H \geq H_{\rm c}$, in the dimer phase, only one gapped mode of $S^{\perp}(\pi,\omega)$ appears in NTENP, while the lower isolated mode of $S^{zz}(\pi,\omega)$ emerges close to $H = H_{\rm c}$ in addition to the isolated $S^{\perp}(\pi,\omega)$ mode. 
As shown so far, the excitation continuum of $S^{zz}(\pi,\omega)$ in the dimer phase shifts to the lower energy region close to $\omega=0$ as compared with that in the Haldane-gap phase. Such different field dependence of the excitation continuum may be intrinsic in the dimer phase and the Haldane-gap phase.

\section{Summary} 
We have investigated the DSF of the $S=1$ bond-alternating Heisenberg chain with a single-ion anisotropy in transverse magnetic fields, using a continued fraction method based on the Lanczos algorithm. 
We have shown that the excitation continuum in $S^{zz}(q,\omega)$ causes characteristic field dependence of the DSF in the Haldane-gap phase and the dimer phase. Our results well reproduce the different field dependence of experimental findings for NTENP and NDMAP.

\section*{Acknowledgments} 
We would like to thank Professor M. Hagiwara, Dr. L. P. Regnault, and Dr. Zheludev for useful comments and valuable discussions. 
Part of our computational programs are based on TITPACK version 2 by H. Nishimori. Numerical computations were carried out at the Yukawa Institute Computer Facility, Kyoto University, and the Supercomputer Center, the Institute for Solid State Physics, University of Tokyo. 
This work was supported by a Grant-in-Aid for Scientific Research from the Ministry of Education, Culture, Sports, Science, and Technology, Japan.

\vspace*{3mm}
\noindent
{\it Note added.}-During the completion of this manuscript, we were informed that in NTENP at $H=0$ the ratio of the lowest-lying energy in $S^{zz}(\pi,\omega)$ to that in $S^{\perp}(\pi,\omega)$ is described by the sine-Gordon quantum field thory \cite{Oshikawa}. The result agrees well with our numerical result.


\begin{references}
%
\bibitem{ah}
I. Affleck and F. D. M. Haldane, Phys. Rev. B {\bf 36}, 5291 (1987).
%
\bibitem{Hagi3}
M. Hagiwara, Y. Narumi, K. Kindo, M. Kohno, H. Nakano, R. Sato, and M. Takahashi, Phys. Rev. Lett. {\bf 80}, 1312 (1998).
%
\bibitem{kohno}
M. Kohno, M. Takahashi, and M. Hagiwara, Phys. Rev. B {\bf 57}, 1046 (1998).
%
\bibitem{nakano}
H. Nakano, M. Hagiwara, T. Chihara, and M. Takahashi, J. Phys. Soc. Jpn. {\bf 66}, 2997 (1997).
%
\bibitem{narumi1}
Y. Narumi, M. Hagiwara, M. Kohno, and K. Kindo, Phys. Rev. Lett. {\bf 86}, 324 (2001).
%
\bibitem{tate}
N. Tateiwa, M. Hagiwara, H. Aruga-Katori, and T. C. Kobayashi, Physica B {\bf 329-333}, 1209 (2003).
%
\bibitem{narumi2}
Y. Narumi, K. Kindo, M. Hagiwara, H. Nakano, A. Kawaguchi, K. Okunishi, and M. Kohno, Phys. Rev. B {\bf 69}, 174405 (2004).
%
\bibitem{yamamoto}
S. Yamamoto, Phys. Rev. B {\bf 51}, 16128 (1995).
%
\bibitem{totsuka}
K.Totsuka, Y. Nishiyama, N. Hatano, and, M. Suzuki, J. Phys.: Condens. Matter {\bf 7}, 4895 (1995).
%
\bibitem{suzuki}
T. Suzuki and S. Suga, Phys. Rev. B {\bf 68}, 134429 (2003). 
%
\bibitem{Regnault}
L. P. Regnault, I. Zaliznyak, J. P. Renard, and C. Vettier, Phys. Rev. B {\bf 50}, 9174 (1994).
%
\bibitem{Tasaki}
H. Tasaki, J. Phys.: Condens. Matter {\bf 3}, 5875 (1991).
%
\bibitem{Hag1}
M. Hagiwara, L. P. Regnault, A. Zheludev, A. Stunault, N. Metoki, T. Suzuki, S. Suga, K. Kakurai, Y. Koike, P. Vorderwisch, and J. H. Chung, to appear in Phys. Rev. Lett.  {\bf 94}, 177202 (2005); cond-mat/0501207.
%
\bibitem{Zhe1}
A. Zheludev, Z. Honda, C. L. Broholm, K. Katsumata, S. M. Shapiro, A. Kolezhuk, S. Park and Y. Qiu, Phys. Rev. B {\bf 68}, 134438 (2003).
%
\bibitem{GB}
E. R. Gagliano and C. A. Balseiro, Phys. Rev. Lett. {\bf 59}, 2999 (1987).  
%
\bibitem{Tak1}
M. Takahashi, Phys. Rev. B {\bf 50}, 3045 (1994).
%
\bibitem{Ma}
S. Ma, C. Broholm, D. H. Reich, B. J. Sternlieb, and R. W. Erwin,  Phys. Rev. Lett. {\bf 69}, 3571 (1992). 
%
\bibitem{Zhe2}
A. Zheludev, Y. Chen, C. L. Broholm, Z. Honda, and K. Katsumata, Phys. Rev. B {\bf 63}, 104410 (2001).  
%
\bibitem{narumi3}
Y. Narumi, M. Hagiwara, R. Sato, K. Kindo, H. Nakano, and M. Takahashi, Physica  B {\bf 246-247}, 509 (1998).
%
\bibitem{tone}
T. Tonegawa, T. Nakano, and M. Kaburagi, J. Phys. Soc. Jpn. {\bf 65}, 3317 (1996); 
W. Chen, K. Hida and C. Sanctuary, J. Phys. Soc. Jpn. {\bf 69}, 237 (2000); 
A. Koga, Phys. Lett. A {\bf 296}, 243 (2002). 
%
\bibitem{com}
In Ref. 15, the phase diagram is represented on the basis of the model Hamiltonian; $
\mathcal{H} = J\sum_{i}[1-\delta (-1)^i]\mathbf{S}_{i} \cdot \mathbf{S}_{i+1}
 + \tilde{D} \sum_{i} (S^{z}_{i})^{2}. 
$
To see the position on this phase diagram, therefore, the parameters have to be described by $(\delta,\tilde{D})$. The parameter sets $(\alpha,D)=(1.0,0.25), (0.8,0.2), (0.45,0.25)$, and $(0.25, 0.08)$ used in this paper are converted into $(\delta,\tilde{D})=(0,0.25), (0.11,0.22), (0.38,0.34)$, and $(0.6,0.13)$, respectively. The first two parameter sets are in the Haldane-gap phase and the latter two ones are in the dimer phase. 
%
\bibitem{fss}
J. L. Cardy, Nucl. Phys. B {\bf 270}, 186 (1986); 
H. W. J. Bl\"{o}te, J. L. Cardy, and M. P. Nightingale, Phys. Rev. Lett. {\bf 56}, 742 (1986); 
I. Affleck, Phys. Rev. Lett. {\bf 56}, 746 (1986).
%
\bibitem{honda}
Z. Honda, H. Asakawa, and K. Katsumata, Phys. Rev. Lett. {\bf 81}, 2566 (1998).
%
\bibitem{Zhe3}
A. Zheludev, T. Masuda, B. Sales, D. Mandrus, T. Papenbrock, T. Barnes, and S. Park, Phys. Rev. B {\bf 69}, 144417 (2004). 
%
\bibitem{Oshikawa}
J. Tamaki and M. Oshikawa, the Physical Society of Japan 2005 Spring Meeting.
%
\end{references}

%

\end{document}